\documentclass[longauth]{aa}

\usepackage{graphicx}
\usepackage{txfonts}
\usepackage{xcolor}
\usepackage{hyperref}
\hypersetup{colorlinks, linkcolor=blue, citecolor=blue, urlcolor=blue}
\usepackage{amsmath}
\usepackage{ulem}
\usepackage{soul}
\usepackage{enumitem}

\newcommand{\pivec}{\mbox{\boldmath $\pi$}}
\newcommand{\muvec}{\mbox{\boldmath $\mu$}}

\newcommand{\te}{t_{\rm E}}
\newcommand{\thetae}{\theta_{\rm E}}

\newcommand{\pie}{\pi_{\rm E}}
\newcommand{\pien}{\pi_{{\rm E},N}}
\newcommand{\piee}{\pi_{{\rm E},E}}
\newcommand{\dl}{D_{\rm L}}
\newcommand{\ds}{D_{\rm S}}

\definecolor{brown}{rgb}{0.59, 0.29, 0.0}
\definecolor{darkgreen}{rgb}{0.0, 0.42, 0.24}
\definecolor{darkblue}{rgb}{0.01, 0.31, 0.59}
\definecolor{darkblue}{rgb}{0.0, 0.25, 0.42}
\definecolor{blue}{rgb}{0.0,0.0,1.0}
\definecolor{green}{rgb}{0.0,1.0,0.0}

\def\eqalign#1{\null\,\vcenter{\openup\jot
        \ialign{\strut\hfil$\displaystyle{##}$&$
        \displaystyle{{}##}$\hfil \crcr#1\crcr}}\,}

\begin{document} 

\title{KMT-2021-BLG-0322: Severe degeneracy between triple-lens and higher-order binary-lens interpretations} 

\author{
     Cheongho Han\inst{\ref{A01}} 
\and Andrew Gould\inst{\ref{A02},\ref{A03}} 
\and Yuki~Hirao\inst{\ref{A04}}
\and Chung-Uk~Lee\inst{\ref{A05}} 
\\                                                
(Leading authors)\\                                 
     Michael~D.~Albrow\inst{\ref{A06}} 
\and Sun-Ju~Chung\inst{\ref{A05}} 
\and Kyu-Ha~Hwang\inst{\ref{A05}} 
\and Youn~Kil~Jung\inst{\ref{A05}} 
\and Doeon~Kim\inst{\ref{A01}} 
\and Shude Mao\inst{\ref{A07},\ref{A08}}
\and Yoon-Hyun~Ryu\inst{\ref{A05}} 
\and In-Gu~Shin\inst{\ref{A05}} 
\and Yossi~Shvartzvald,\inst{\ref{A09}} 
\and Jennifer~C.~Yee\inst{\ref{A10}} 
\and Weicheng Zang\inst{\ref{A07}} 
\and Sang-Mok~Cha\inst{\ref{A05},\ref{A11}} 
\and Dong-Jin~Kim\inst{\ref{A05}} 
\and Hyoun-Woo~Kim\inst{\ref{A05}} 
\and Seung-Lee~Kim\inst{\ref{A05}} 
\and Dong-Joo~Lee\inst{\ref{A05}}
\and Yongseok~Lee\inst{\ref{A05}} 
\and Byeong-Gon~Park\inst{\ref{A05}} 
\and Richard~W.~Pogge\inst{\ref{A02}}
\\
(The KMTNet Collaboration),\\
     Fumio~Abe\inst{\ref{A12}}                           
\and Richard Barry\inst{\ref{A13}}                       
\and David~P.~Bennett\inst{\ref{A13},\ref{A14}}          
\and Aparna~Bhattacharya\inst{\ref{A13},\ref{A14}}       
\and Ian Bond\inst{\ref{A15}}                            
\and Martin~Donachie\inst{\ref{A15}}                     
\and Hirosane~Fujii\inst{\ref{A04}}                     
\and Akihiko~Fukui\inst{\ref{A16},\ref{A17}}            
\and Yoshitaka~Itow\inst{\ref{A12}}                     
\and Rintaro~Kirikawa\inst{\ref{A04}}                   
\and Iona~Kondo\inst{\ref{A04}}                         
\and Naoki~Koshimoto\inst{\ref{A18},\ref{A19}}           
\and Man~Cheung~Alex~Li\inst{\ref{A20}}                 
\and Yutaka~Matsubara\inst{\ref{A12}}                   
\and Yasushi~Muraki\inst{\ref{A12}}                     
\and Shota~Miyazaki\inst{\ref{A04}}                      
\and Cl\'ement~Ranc\inst{\ref{A13}}                      
\and Nicholas~J.~Rattenbury\inst{\ref{A20}}              
\and Yuki~Satoh\inst{\ref{A04}}                         
\and Hikaru~Shoji\inst{\ref{A04}}                       
\and Takahiro~Sumi\inst{\ref{A04}}                      
\and Daisuke~Suzuki\inst{\ref{A21}}
\and Yuzuru~Tanaka\inst{\ref{A04}}
\and Paul~J.~Tristram\inst{\ref{A22}}
\and Tsubasa~Yamawaki\inst{\ref{A04}}
\and Atsunori~Yonehara \inst{\ref{A04}}
\\
(The MOA Collaboration)\\ 
}

\institute{
       Department of Physics, Chungbuk National University, Cheongju 28644, Republic of Korea  \\ \email{cheongho@astroph.chungbuk.ac.kr}                 \label{A01}
\and   Department of Astronomy, The Ohio State University, 140 W.  18th Ave., Columbus, OH 43210, USA                                                     \label{A02}
\and   Max Planck Institute for Astronomy, K\"onigstuhl 17, D-69117 Heidelberg, Germany                                                                   \label{A03}
\and   Department of Earth and Space Science, Graduate School of Science, Osaka University, Toyonaka, Osaka 560-0043, Japan                               \label{A04}
\and   Korea Astronomy and Space Science Institute, Daejon 34055,Republic of Korea                                                                        \label{A05}
\and   University of Canterbury, Department of Physics and Astronomy, Private Bag 4800, Christchurch 8020, New Zealand                                    \label{A06}
\and   National Astronomical Observatories, Chinese Academy of Sciences, Beijing 100101, China                                                            \label{A07}
\and   Department of Astronomy, Tsinghua University, Beijing 100084, China                                                                                \label{A08}
\and   Department of Particle Physics and Astrophysics, WeizmannInstitute of Science, Rehovot 76100, Israel                                               \label{A09}
\and   Center for Astrophysics|Harvard \& Smithsonian 60 Garden St., Cambridge, MA 02138, USA                                                             \label{A10}
\and   School of Space Research, Kyung Hee University, Yongin, Kyeonggi 17104, Republic of Korea                                                          \label{A11}
\and   Institute for Space-Earth Environmental Research, Nagoya University, Nagoya 464-8601, Japan                                                        \label{A12}
\and   Code 667, NASA Goddard Space Flight Center, Greenbelt, MD 20771, USA                                                                               \label{A13}
\and   Department of Astronomy, University of Maryland, College Park, MD 20742, USA                                                                       \label{A14} 
\and   Institute of Natural and Mathematical Sciences, Massey University, Auckland 0745, New Zealand                                                      \label{A15}
\and   Department of Earth and Planetary Science, Graduate School of Science, The University of Tokyo, 7-3-1 Hongo, Bunkyo-ku, Tokyo 113-0033, Japan      \label{A16}
\and   Instituto de Astrof\'isica de Canarias, V\'ia L\'actea s/n, E-38205 La Laguna, Tenerife, Spain                                                     \label{A17}
\and   Department of Astronomy, Graduate School of Science, The University of Tokyo, 7-3-1 Hongo, Bunkyo-ku, Tokyo 113-0033, Japan                        \label{A18}
\and   National Astronomical Observatory of Japan, 2-21-1 Osawa, Mitaka, Tokyo 181-8588, Japan                                                            \label{A19}
\and   Department of Physics, University of Auckland, Private Bag 92019, Auckland, New Zealand                                                            \label{A20}
\and   Institute of Space and Astronautical Science, Japan Aerospace Exploration Agency, 3-1-1 Yoshinodai, Chuo, Sagamihara, Kanagawa, 252-5210, Japan    \label{A21}
\and   University of Canterbury Mt.\ John Observatory, P.O.\ Box 56, Lake Tekapo 8770, New Zealand                                                        \label{A22}
}
\date{Received ; accepted}

\abstract
{}
{
We investigate the microlensing event KMT-2021-BLG-0322, for which the light curve 
exhibits three distinctive sets of caustic-crossing features. It is found that the overall 
features of the light curve are approximately described by a binary-lens (2L1S) model, but 
the model leaves substantial residuals. We test various interpretations with the aim of 
explaining the residuals.
}
{
We find that the residuals can be explained either by considering a nonrectilinear 
lens-source motion caused by the microlens-parallax and lens-orbital effects or by adding 
a low-mass companion to the binary lens (3L1S model). The degeneracy between the higher-order 
2L1S model and the 3L1S model is very severe, making it difficult to single out a correct 
solution based on the photometric data. This degeneracy was known before for two previous 
events (MACHO-97-BLG-41 and OGLE-2013-BLG-0723), which led to the false detections of planets 
in binary systems, and thus the identification of the degeneracy for KMT-2021-BLG-0322
illustrates that the degeneracy can be not only common but also very severe, emphasizing the 
need to check both interpretations of deviations from 2L1S models.  
}
{
From the Bayesian analysis conducted with the measured lensing observables of the event 
timescale, angular Einstein radius, and microlens parallax, it was estimated that the 
binary lens components have masses 
$(M_1, M_2) =(0.62^{+0.25}_{-0.26}~M_\odot,0.07^{+0.03}_{-0.03}~M_\odot)$, 
for both 2L1S and 3L1S solutions, and the mass of the tertiary lens component according to 
the 3L1S solution is $M_3=6.40^{+2.64}_{-2.78}~M_{\rm J}$.

}
{}


\maketitle

\section{Introduction}\label{sec:one}

Searching for planets in binary or multiple systems is of scientific importance for two major
reasons. First, planets in multiple systems act as test beds for investigating various mechanisms 
of planet formation and evolution because the gravity of the stellar companion may influence the 
formation and the subsequent dynamical evolution of planets. Second, systematic searches for planets 
in multiple systems are important in the estimation of the global planet frequency because binaries 
are very common among field stars.  An up-to-date overview of exoplanet statistics and theoretical 
implications is available in \citet{Zhu2021}.

Gravitational microlensing is an important tool for detecting planets in binary systems, 
especially those belonging to binaries composed of faint stars, which are the most common 
population of stars in the Galaxy.  Microlensing detections of planets in such systems are 
possible because of the lensing characteristic that does not depend on the luminosity of a 
lensing object. There exist six confirmed microlensing planets in binary systems, including 
OGLE-2008-BLG-092L \citep{Poleski2014}, OGLE-2007-BLG-349L \citep{Bennett2016}, OGLE-2013-BLG-0341L 
\citep{Gould2014}, OGLE-2016-BLG-0613L \citep{Han2017}, OGLE-2018-BLG-1700L \citep{Han2020}, and 
KMT-2020-BLG-0414L \citep{Zang2021}, and for all of these systems, the planet hosts are less 
massive, and thus fainter, than the Sun. Besides these systems, the lens of the event 
OGLE-2019-BLG-0304 is likely to be a planetary system in a binary, but this interpretation is 
not conclusive due to the possibility of an alternative interpretation \citep{Han2021}.

Planets in binaries manifest themselves via signals of various types in lensing light curves. 
The first type is an independent short single-lensing (1L1S) light curve that is well separated 
from the binary-lensing (2L1S) light curve produced by the host binary stars. The planetary 
signals in the lensing events OGLE-2008-BLG-092 and OGLE-2013-BLG-0341 were detected through this 
channel. The signal of the second type is generated by the source passage through the mingled caustic 
region, in which the two sets of lensing caustics induced by the planet and binary companion overlap. 
Here the caustic refers to the positions on the source plane at which the lensing magnification 
of a point source would be infinite.  In this case, the planetary signal superposes with the signal 
of the binary companion \citep{Lee2008}. The planetary signals in the lensing events OGLE-2016-BLG-0613 and 
OGLE-2018-BLG-1700 were found through this channel.  The signal of the third type is a small distortion 
from a 2L1S lensing light curve caused by the presence of a tertiary lens component, either a companion 
binary star or a planet. Such a signal was detected in the case of the lensing event OGLE-2007-BLG-349, 
for which the overall lensing light curve was approximated as that of a 2L1S event produced by a 
star-planet pair, but an additional binary companion to the host was needed to precisely describe the 
observed light curve.  In addition, although the major planetary signal of OGLE-2013-BLG-0341 was detected 
through the independent channel, the presence of the planet was additionally confirmed by the deviation of 
the 2L1S model in the region of the binary-induced anomaly.

Light curves produced by triple-lens (3L1S) systems with planets can be degenerate with those 
of 2L1S events deformed by higher-order effects.  This degeneracy was known for two previous 
events MACHO-97-BLG-41 and OGLE-2013-BLG-0723.  For both events, the light curves were originally 
interpreted as 3L1S events by \citet{Bennett1999} and \citet{Udalski2015}, respectively, and it 
was subsequently shown that the signals of the third body were spurious and the residuals from 
the 2L1S models could be explained with the consideration of the lens orbital motion by 
\citet{Albrow2000} and \citet{Jung2013} for MACHO-97-BLG-41 and by \citet{Han2016} for 
OGLE-2013-BLG-0723.

In this work, we report the analysis of the lensing event KMT-2021-BLG-0322/MOA-2021-BLG-091.
The light curve of the event exhibits three distinctive sets of caustic features, for all of 
which the rising and falling sides of the caustic crossings were resolved by lensing surveys. 
Although the overall features of the lensing light curve are approximately described by a 2L1S 
model, the model leaves residuals from the model. We investigate the origin of the residuals by 
testing various interpretations of the lensing system.

For the presentation of the analysis, we organize the paper as follows. In Sect.~\ref{sec:two}, 
we describe observations conducted to acquire the data used in the analysis. The anomalous 
features in the observed lensing light curve are depicted in the same section. In Sect.~\ref{sec:three}, 
we conduct modeling of the observed lensing light curve under two interpretations of the lens 
system and present the results of the analysis. In Sect.~\ref{sec:four}, we describe the detailed 
procedure of measuring the lensing observables that can constrain the physical lens parameters. 
In Sect.~\ref{sec:five}, we estimate the physical lens parameters. In Sect.~\ref{sec:six}, we 
summarize the result and conclude.

\begin{figure}[t]
\includegraphics[width=\columnwidth]{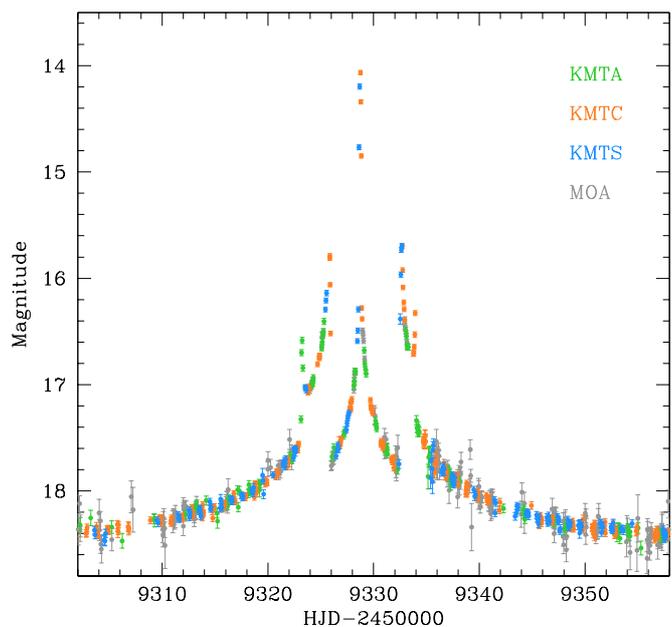}
\caption{
Light curve of the lensing event KMT-2021-BLG-0322.  The colors of the data points are set to 
match those of the telescopes used for the data acquisition.
}
\label{fig:one}
\end{figure}

\section{Observations and data}\label{sec:two}

The source star of the lensing event KMT-2021-BLG-0322/MOA-2021-BLG-091 is located toward the Galactic 
bulge field at the equatorial coordinates $({\rm RA}, {\rm decl.})=(18:03:38.80, -29:36:15.80)$, which 
correspond to the galactic coordinates $(l, b) = (1^\circ\hskip-2pt .409, -3^\circ\hskip-2pt .731)$. 
The source brightness had remained constant before the lensing magnification with a baseline magnitude 
of $I_{\rm base}=18.49$ according to the KMTNet scale.

The increase in the source flux induced by lensing was first found by the AlertFinder algorithm
\citep{Kim2018} of the Korea Microlensing Telescope Network \citep[KMTNet:][]{Kim2016} survey
on 2021-04-09 (${\rm HJD}^\prime \equiv {\rm HJD}-2450000\sim 9313.5$), when the source became 
brighter than the baseline by $\Delta I\sim0.34$~mag. The Microlensing Observations in Astrophysics 
\citep[MOA:][]{Bond2001} group independently found the event 11 days after the KMTNet discovery and 
dubbed the event as MOA-2021-BLG-091. Hereafter, we designate the event as KMT-2021-BLG-0322 following 
the convention that the event ID of the first discovery survey is used as a representative designation.

The KMTNet observations of the event were conducted utilizing the three 1.6~m telescopes located in 
the three continents of the Southern Hemisphere: the Siding Spring Observatory in Australia (KMTA), 
the Cerro Tololo Interamerican Observatory in Chile (KMTC), and the South African Astronomical 
Observatory in South Africa (KMTS).  Each KMTNet telescope is equipped with a camera yielding 
4~deg$^2$ field of view. The MOA survey used the 1.8~m telescope of the Mt.~John Observatory in 
New Zealand, and the camera mounted on the telescope yields a 2.2 deg$^2$ field of view.  The 
principal observations of the KMTNet and MOA surveys were done in the $I$ band and the customized 
MOA-Red band, respectively, and for both surveys, a fraction of images were acquired in the $V$ 
band to measure the color of the source star.  In Sect.~\ref{sec:four}, we present the procedure 
of the source color measurement.

\begin{figure}[t]
\includegraphics[width=\columnwidth]{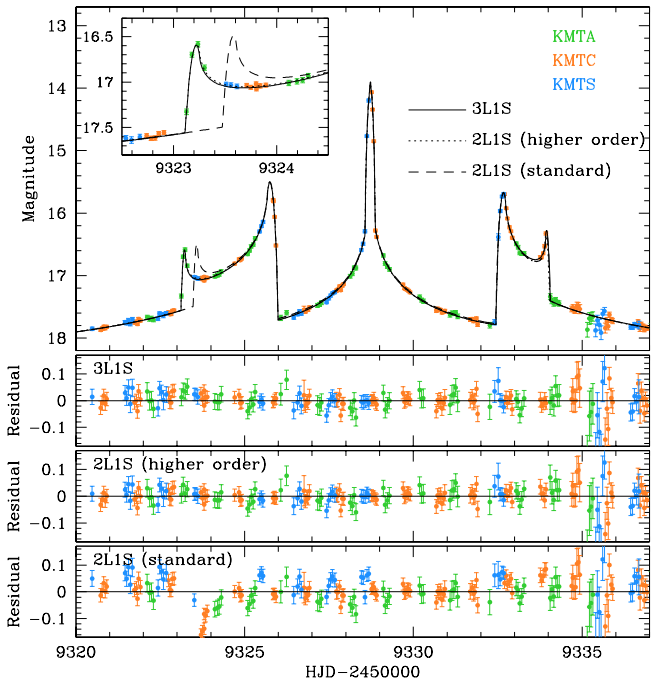}
\caption{
Zoom-in view of the anomaly region of the lensing light curve.  The inset in the top panel 
shows the enlarged view around the first peak.  The solid, dotted, and dashed curves drawn over 
the data points are the model curves of the 3L1S, higher-order 2L1S, and standard 2L1S 
solutions, respectively.  The three lower panels show the residuals from the individual
models.  The curves of the 3L1S and higher-order 2L1S models are difficult to be distinguished
 with the line width.
}
\label{fig:two}
\end{figure}

In Figure~\ref{fig:one}, we present the lensing light curve of the event constructed using the 
combined data from the KMTNet and MOA surveys. The light curve is characterized by five strong 
peaks that occurred at ${\rm HJD}^\prime \sim 9323.1$, 9321.9, 9329.7, 9332.4, and 9334.0. 
Figure~\ref{fig:two} shows the zoom-in view of the peaks. The U-shape trough region between the 
first and second peaks suggests that these peaks are a pair of spikes produced when the source 
entered and exited a caustic. A similar pattern between the fourth and fifth peaks suggests the 
same origin as those of the first and second peaks.  The third peak does not exhibit a U-shape 
trough, and this suggests that the peak was produced by the source passage over the tip (cusp) 
of a caustic.  We note that all the caustic spikes were resolved by the combination of the KMTNet 
data sets, which were acquired with a 1.0~hr cadence for KMTC data set and  with a 0.75~hr cadence 
for KMTA+KMTS data sets.  Although the event was not in the KMTNet prime fields, which are  
covered with a 0.25 hr cadence, the cadence of the KMTNet observations was adequate in part because 
of relatively long durations of the caustic crossings.  In the analysis, we did not use the MOA data 
first because these data did not resolve any of the caustic spikes, and second because the precision 
of the data was low compared to the KMTNet data sets, especially at lower magnifications.

Data reduction was carried out using the KMTNet photometry pipeline \citep{Albrow2017} developed 
based on the difference imaging algorithm \citep{Tomaney1996, Alard1998}.  Following the standard 
procedure described in \citet{Yee2012}, we rescaled the error bars of the data estimated from the 
pipeline, $\sigma_0$, by 
\begin{equation}
\sigma=k(\sigma_{\rm min}^2+\sigma_0^2)^{1/2}, 
\label{eq1}
\end{equation}
where $\sigma_{\rm min}$ is a factor added in the quadrature to take into account the scatter of 
data, and the other factor $k$ is used to make $\chi^2$ per degree of freedom for each data set 
unity. In Table~\ref{table:one}, we list the values of $k$ and $\sigma_{\rm min}$ along with the 
numbers of data points, $N_{\rm data}$, for the individual data sets.

For the event, the light curve was analyzed in real time with the progress of the event. Y.~Hirao
of the MOA group first released a lensing model when the data covered the first two peaks, and this 
model interpreted the event as a 2L1S events produced by a binary lens with a projected separation 
(normalized to the angular Einstein radius $\thetae$) and mass ratio of $(s, q)\sim (1.1, 0.1)$.  
C.~Han of the KMTNet group conducted modeling after the third peak was covered, and found a model 
that is similar to that of Y.~Hirao.  From the additional modeling conducted with updated data after 
the final peak was covered, C.~Han realized that a 2L1S model under the assumption of a rectilinear 
relative lens-source motion (standard 2L1S model) could not precisely describe all the caustic features, 
although the model described the overall pattern of the light curve.  In the following section, we 
depict in detail the inadequacy of the standard 2L1S model in describing the observed data and 
investigate the origin of this deviation.

\section{Interpretations of lensing light curve}\label{sec:three}

\subsection{2L1S interpretation}\label{sec:three-one}

The caustic features in the lensing light curve suggest that the event was produced by a lens 
composed of multiple masses.  We, therefore, start with a standard 2L1S model for the interpretation 
of the observed lensing light curve.  The modeling is done by searching for the set of the lensing 
parameters that best describe the observed light curve. For a 1L1S event, the lensing light curve 
is described by three parameters of $(t_0, u_0, \te)$, which are the time of the closest lens-source 
approach, the separation at that time, and the event timescale, respectively.  Describing a 
caustic-crossing 2L1S light curve requires four additional parameters of $(s, q, \alpha, \rho)$, 
which denote the separation and mass ratio between the binary lens components, $M_1$ and $M_2$, the 
angle between the source trajectory and the line connecting the binary lens components, and the ratio 
of the angular source radius $\theta_*$ to the angular Einstein radius, that is, $\rho=\theta_*/\thetae$ 
(normalized source radius), respectively. The normalized source radius is included in modeling to account 
for finite-source effects during the caustic crossings of the source.

\begin{table}[t]
\small
\caption{Error bar normalization factors\label{table:one}}
\begin{tabular*}{\columnwidth}{@{\extracolsep{\fill}}ccccc}
\hline\hline
\multicolumn{1}{c}{Data set}                  &
\multicolumn{1}{c}{$k$}                       &
\multicolumn{1}{c}{$\sigma_{\rm min}$ (mag)}  &
\multicolumn{1}{c}{$N_{\rm data}$}            \\
\hline
KMTA &  1.135  &  0.02  &  121 \\
KMTC &  1.000  &  0.02  &  345 \\ 
KMTS &  1.072  &  0.02  &  178 \\
\hline
\end{tabular*}
\end{table}

The 2L1S modeling is conducted in two steps. In the first step, we divide the lensing parameters 
into two groups, and the parameters $s$ and $q$ are searched for via a grid approach, while the 
other parameters are found via a downhill approach. For the downhill approach, we use the Markov 
Chain Monte Carlo (MCMC) algorithm. For the MCMC parameters $(t_0, u_0, \te)$, the initial values are 
given based on the peak time, magnification, and duration of the event. For the source trajectory 
angle $\alpha$, modeling is done with multiple starting values that are evenly divided in the 
0--$2\pi$ range. From the modeling in this step, we obtain a $\Delta\chi^2$ map on the $s$--$q$ 
parameter plane, and identify local solutions. In the second step, we refine the local solutions 
identified in the first-round modeling by allowing all lensing parameters to vary.  For the 
computations of finite-source magnifications, we use the map-making method described in 
\citet{Dong2006}.

\begin{figure}[t]
\includegraphics[width=\columnwidth]{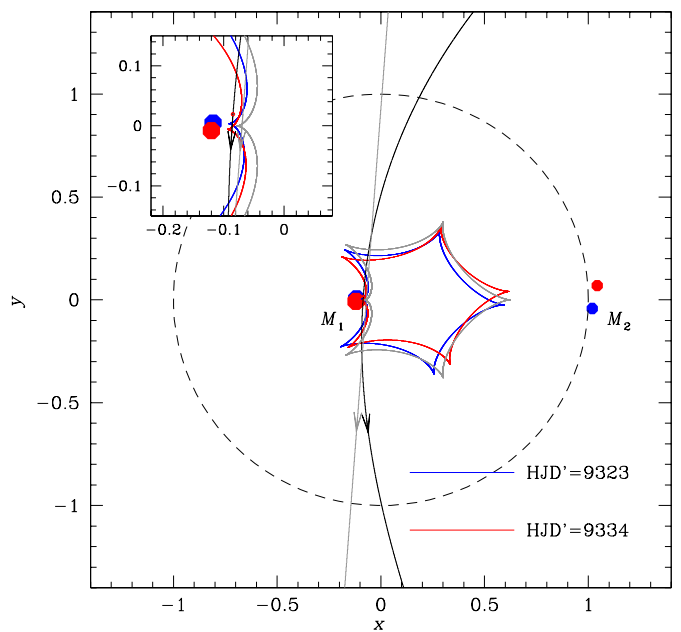}
\caption{
Lens system configuration according to the 2L1S model.  The concave curve represents the caustic, 
the line with an arrow is the source trajectory, and the filled dots marked by $M_1$ and $M_2$ 
indicate the positions of the lens components.  The caustic of the standard model are drawn in 
gray.  For the higher-order model, in which the lens position and caustic vary in time 
because of the lens orbital motion, we mark the lens and caustic at two epochs of ${\rm HJD}^\prime
=9323$ (marked in blue color) and 9334 (in red color).  The source trajectories of the standard and 
higher-order models are drawn in gray and black.  The inset shows the zoom-in view of the 
region around $M_1$.  The coordinates are centered at the photocenter and lengths are scaled to 
the angular Einstein radius corresponding to the total mass of the lens.  The dashed unit circle 
centered at the origin represents the Einstein ring.
}
\label{fig:three}
\end{figure}

In Figure~\ref{fig:two}, we present the 2L1S model curve (dashed curve plotted over the data points).  
The lensing parameters of the model (standard 2L1S model) are listed in Table~\ref{table:two}. 
The binary lensing parameters $(s, q)\sim (1.1, 0.14)$ are similar to those of the Hirao model obtained 
from the modeling conducted in the early stage of the event.  The lens system configuration is shown 
in Figure~\ref{fig:three}, in which the caustic and the source trajectory are drawn in gray. 
The coordinates of the configuration are scaled to $\thetae$ corresponding to the total mass of the 
binary lens, and the origin of the coordinates is set at the photocenter defined by \citet{Stefano1996} 
and \citet{An2002}.  According to the best-fit 2L1S model, the caustic forms a single closed curve 
with six cusps (resonant caustic), and the source passed the left side of the caustic, passing through 
the caustic three times, with the individual passages producing the caustic-crossing features in the 
lensing light curve. Due to the resonant nature of the caustic, the solution is unique without any 
degeneracy.

\begin{table}[t]
\small
\caption{Best-fit lensing parameters of 2L1S solution\label{table:two}}
\begin{tabular*}{\columnwidth}{@{\extracolsep{\fill}}lcccc}
\hline\hline
\multicolumn{1}{c}{Parameter}      &
\multicolumn{1}{c}{Standard}      &
\multicolumn{1}{c}{Higher-order}          \\
\hline
$\chi^2$                       &  3049.1                &           632.3            \\
$t_0$ (HJD$^\prime$)           &  9328.618 $\pm$ 0.001  &  9328.546 $\pm$ 0.009      \\ 
$u_0$                          &  -0.069   $\pm$ 0.001  &    -0.086 $\pm$ 0.002      \\
$\te$ (days)                   &  21.19    $\pm$ 0.26   &    25.27  $\pm$ 0.61       \\
$s$                            &  1.135    $\pm$ 0.002  &     1.148 $\pm$ 0.005      \\
$q$                            &  0.143    $\pm$ 0.002  &     0.115 $\pm$ 0.003      \\ 
$\alpha$ (rad)                 &  4.787    $\pm$ 0.001  &     4.782 $\pm$ 0.008      \\
$\rho$ (10$^{-3}$)             &  3.306    $\pm$ 0.060  &     2.683 $\pm$ 0.073      \\
$\pien$                        &  --                    &    -2.31  $\pm$ 2.05       \\   
$\piee$                        &  --                    &    -0.28  $\pm$ 0.18       \\   
$ds/dt$ (yr$^{-1}$)            &  --                    &     0.93  $\pm$ 0.04       \\   
$d\alpha/dt$ (yr$^{-1}$)       &  --                    &     3.57  $\pm$ 2.12       \\   
\hline                                                     
\end{tabular*}
\tablefoot{ ${\rm HJD}^\prime = {\rm HJD}- 2450000$.  }
\end{table}

It is found that the standard 2L1S model cannot precisely explain the observed light curve. This 
can be seen in the bottom panels of Figure~\ref{fig:two}, in which we present the residuals from 
the 2L1S model.  Although the model depicts the overall features of the light curve, there is an 
offset in the time of the first peak between the model and observed data (see the zoom-in view 
around the fist caustic crossing shown in the inset inserted in the upper panel of Figuree~\ref{fig:two}).  
Besides the region of this peak, there exist noticeable residuals throughout the region of the caustic 
features. This indicates that the standard 2L1S interpretation is not adequate and a more sophisticated 
model is needed for the precise description of the data.

We checked the possibility that the major deviation from the 2L1S model, that is, the offset 
between the model and data in the region of the first peak, can be explained by higher-order 
effects.  We considered two higher-order effects that cause the relative lens-source motion to 
be nonrectilinear: microlens-parallax and lens-orbital effects. The former effect is caused by 
the nonlinear motion of an observer caused by the Earth's orbital motion around the Sun 
\citep{Gould1992}, and the latter effect is caused by the orbital motion of the binary lens 
\citep{Dominik1998, Ioka1999}. In order to check these higher-order effects, we conduct 
additional modeling by adding two extra pairs of parameters $(\pien, \piee)$ and $(ds/dt, 
d\alpha/dt)$: higher-order 2L1S model.  The first pair of the higher-order lensing parameters 
represent the north and east components of the microlens-parallax vector 
$\pivec_{\rm E}$, 
\begin{equation}
\pivec_{\rm E} \equiv {\pi_{\rm rel}\over\thetae}\,{\muvec_{\rm rel}\over \mu_{\rm rel}},
\label{eq2}
\end{equation}
respectively, and the second pair represent the annual change rates of the binary separation 
and source trajectory angle, respectively.  Here $\muvec_{\rm rel}$ represents the relative 
lens-source proper motion, $\pi_{\rm rel}= {\rm AU}(D_{\rm S}^{-1}-D_{\rm L}^{-1})$ is the 
relative lens-source parallax, and $\dl$ and $\ds$ denote the distances to the lens and source, 
respectively.  In this modeling, we restrict the lensing parameters to satisfy the dynamical 
condition of $ ({\rm KE/PE})_\perp \leq {\rm KE/PE} \leq 1.0$, where $({\rm KE/PE})$ and 
$({\rm KE/PE})_\perp$ represent the intrinsic and projected kinetic-to-potential energy ratio, 
respectively \citep{Dong2009}.

\begin{figure}[t]
\includegraphics[width=\columnwidth]{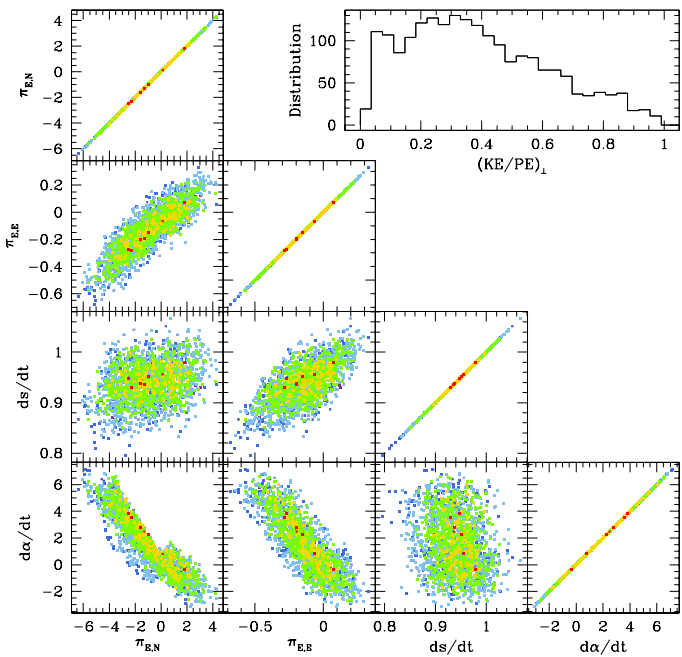}
\caption{
$\Delta\chi^2$ distribution of points in the MCMC chain of the 2L1S solution on the planes 
of higher-order parameters: $\pien$, $\piee$, $ds/dt$, and $d\alpha/dt$.
Points with different colors indicate those with $\Delta\chi^2\leq 1\sigma$ (red), $\leq 2\sigma$ 
(yellow), $\leq 3\sigma$ (green), $\leq 4\sigma$ (cyan), and $\leq 5\sigma$ (blue).
The upper right panel shows the distribution of the projected kinetic-to-potential energy ratio.
}
\label{fig:four}
\end{figure}

We find that the residuals from the standard 2L1S model are greatly reduced with the 
consideration of the higher-order effects.  This is shown in Figure~\ref{fig:two}, in 
which we present the model curve (dotted curve) and residuals from the model.  It is found 
that the higher-order model explains not only the major deviation in the offset of the 
first peak but also improves the fit throughout the anomaly region.  The $\chi^2$ difference 
between the standard and higher-order models is $\Delta\chi^2=2416.8$.  The lensing parameters 
of the higher-order 2L1S model are listed in Table~\ref{table:two}.  The lens system configuration 
of the higher-order 2L1S model is presented in Figure~\ref{fig:three}, which shows that  the 
caustic shape varies in time due to the lens-orbital effect and that the source trajectory 
becomes nonrectilinear due to the microlens-parallax effect.  Figure~\ref{fig:four} shows the 
$\Delta\chi^2$ distribution of points in the MCMC chain on the planes of higher-order parameters.  
It shows that these parameters are strongly correlated because the microlens-parallax and 
lens-orbital effects induce similar deviations in lensing light curves \citep{Batista2011}.  
As a result, the uncertainties of the higher-order lensing parameters are considerable, for example, 
$\pien=-2.31 \pm 2.05$ and $d\alpha/dt=(3.57 \pm 2.12)$~yr$^{-1}$, despite the obvious effects on 
the lensing light curve.  Also shown in Figure~\ref{fig:four} is the distribution of the projected 
kinetic-to-potential energy ratio.  The mean and standard deviation of the ratio distribution are 
$({\rm KE/PE})_\perp =0.46\pm 0.25$, which is well within the range of 
$0.2\lesssim ({\rm KE/PE})_\perp\lesssim 0.5$ for moderate eccentricity binaries that are observed 
at usual viewing angles.

\begin{table*}[t]
\small
\caption{Best-fit lensing parameters of 3L1S solutions\label{table:three}}
\begin{tabular}{lcccc}
\hline\hline
\multicolumn{1}{c}{Parameter}                &
\multicolumn{2}{c}{Wide}                     &
\multicolumn{2}{c}{Close}                    \\
\multicolumn{1}{c}{}                         &
\multicolumn{1}{c}{standard}                 &
\multicolumn{1}{c}{parallax}                 &
\multicolumn{1}{c}{standard}                 &
\multicolumn{1}{c}{parallax}                 \\
\hline
$\chi^2$             &  642.1                 &  630.5                     &  665.2                 &  642.3                        \\
$t_0$ (HJD$^\prime$) &  9328.686 $\pm$ 0.003  &  9328.682 $\pm$ 0.005      &  9328.661 $\pm$ 0.002  &  9328.674 $\pm$  0.006        \\ 
$u_0$                &  -0.053   $\pm$ 0.001  &    -0.050 $\pm$ 0.002      &  -0.057   $\pm$ 0.001  &    -0.053 $\pm$  0.002        \\
$\te$ (days)         &  25.04    $\pm$ 0.50   &    26.37  $\pm$ 0.62       &  24.99    $\pm$ 0.53   &    25.88  $\pm$  0.54         \\
$s_2$                &  1.124    $\pm$ 0.003  &     1.135 $\pm$ 0.006      &  1.129    $\pm$ 0.003  &     1.139 $\pm$  0.006        \\
$q_2$                &  0.109    $\pm$ 0.003  &     0.104 $\pm$ 0.003      &  0.115    $\pm$ 0.002  &     0.109 $\pm$  0.004        \\ 
$\alpha$ (rad)       &  4.762    $\pm$ 0.001  &     4.765 $\pm$ 0.002      &  4.768    $\pm$ 0.001  &     4.766 $\pm$  0.003        \\
$s_3$                &  1.231    $\pm$ 0.029  &     1.239 $\pm$ 0.040      &  0.788    $\pm$ 0.009  &     0.711 $\pm$  0.011        \\
$q_3$ ($10^{-3}$)    &  9.88     $\pm$ 0.90   &     9.75  $\pm$ 1.32       &  3.20     $\pm$ 0.20   &     5.97  $\pm$  0.64         \\ 
$\psi$ (rad)         &  0.321    $\pm$ 0.017  &     0.295 $\pm$ 0.022      &  0.519    $\pm$ 0.003  &     0.431 $\pm$  0.009        \\
$\rho$ (10$^{-3}$)   &  2.76     $\pm$ 0.07   &     2.66  $\pm$ 0.08       &  2.76     $\pm$ 0.07   &     2.68  $\pm$  0.08         \\
$\pien$              &  --                    &    -0.06  $\pm$ 0.91       &  --                    &    -0.14  $\pm$  1.20         \\  
$\piee$              &  --                    &    -0.30  $\pm$ 0.091      &  --                    &    -0.43  $\pm$  0.093        \\  
\hline
\end{tabular}
\end{table*}

\subsection{3L1S interpretation}\label{sec:three-two}

We also test a 3L1S modeling for the interpretation of the residuals from the standard 2L1S 
model.  In addition to the 2L1S lensing parameters, a 3L1S modeling requires extra parameters 
in order to describe the third lens component, $M_3$. These parameters are $(s_3, q_3, \psi)$, 
which denote the projected separation and mass ratio between $M_1$ and $M_3$, and the orientation 
angle of $M_3$ as measured from the $M_1$--$M_2$ axis with a center at the position of $M_1$. 
Hereafter, we denote the separation and mass ratio between $M_1$ and $M_2$ as $(s_2, q_2)$ to 
distinguish them from those of the third body.  We check a 3L1S model because it is known that a 
third body of a lens can induce a small distortion of the caustic, particularly in the neighborhood 
of a cusp, and this may explain the deviations from the standard 2L1S model.  For example, 
\citet{Gould2014} showed that the three parameters $(s_3,q_3,\psi)$ of the planetary companion in 
the 3L1S event OGLE-2013-BLG-0341 could be fully recovered even when the data that were affected 
by the planetary caustic were eliminated from the fit, due to the distortion induced by the planet 
on a cusp associated with the binary-lens caustic. In particular, they argued that planets could 
be discovered from such cusp distortions even when the source did not pass near or over the 
planetary caustic.

\begin{figure}[t]
\includegraphics[width=\columnwidth]{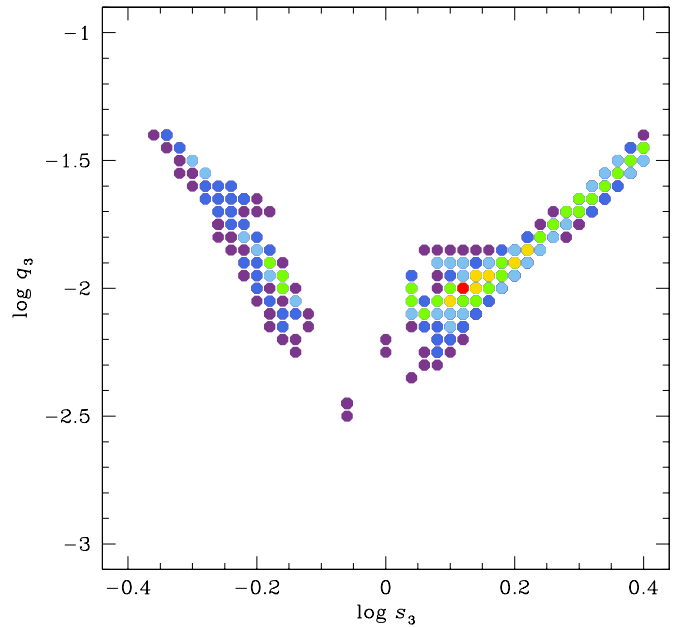}
\caption{
$\Delta\chi^2$ map on the  $\log s_3$--$\log q_3$ plane obtained from the grid search. Color 
coding is set to represent points with $\Delta\chi^2\leq 1n\sigma$ (red), $\leq 2n\sigma$ (yellow), 
$\leq 3n\sigma$ (green), $\leq 4n\sigma$ (cyan), $\leq 5n\sigma$ (blue), $\leq 6n\sigma$ (purple),
where $n=3$.
}
\label{fig:five}
\end{figure}

Similar to the 2L1S modeling, the 3L1S modeling is carried out in two steps. In the first step, 
we conduct grid searches for the triple-lens parameters, that is, $(s_3, q_3, \psi)$, by fixing 
the 2L1S parameters as those of the best-fit standard 2L1S model. We fix the 2L1S parameters because 
the 2L1S model describes the overall pattern of the light curve and the variation in the lensing 
light curve by the third body would be minor \citep{Bozza1999, Han2001}.  This step yields 
$\Delta\chi^2$ maps on the $s_3$--$q_3$-$\psi$ planes, and we identify local solutions from 
the maps. In the second step, the individual local solutions found from the first step are 
refined by allowing all parameters to vary.

\begin{figure}[t]
\includegraphics[width=\columnwidth]{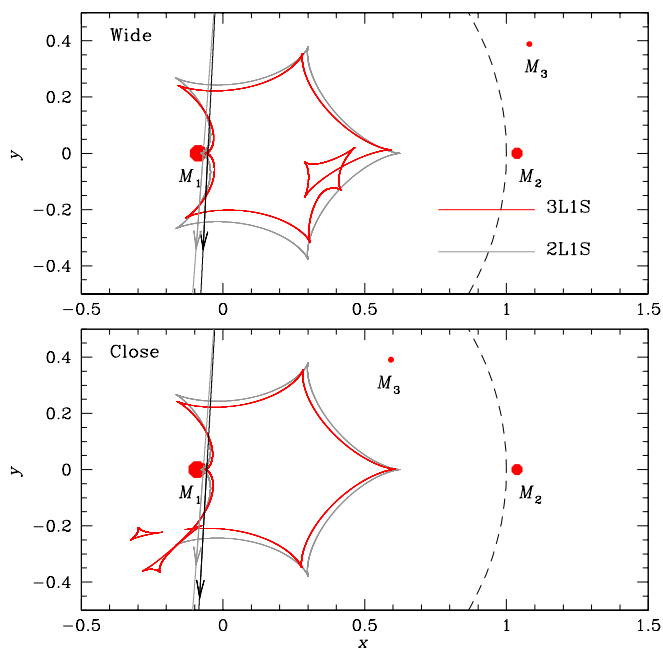}
\caption{
Lens system configurations according to the wide (upper panel) and close (lower panel) 3L1S 
solutions. In each panel, the nested concave curve represents the caustic, the line with an 
arrow is the source trajectory, and the three filled red dots marked by $M_1$, $M_2$, and 
$M_3$ indicate the positions of the lens components. The caustic curve and the source 
trajectory according to the static 2L1S model are drawn in gray to show the caustic 
variation by the third body $M_3$.  Other notations are same as those in Fig.~\ref{fig:three}.
}
\label{fig:six}
\end{figure}

Figure~\ref{fig:five} shows the $\Delta\chi^2$ map on the $\log s_3$--$\log q_3$ plane obtained 
from the grid search. The map shows two locals at $(\log s_3, \log q_3)\sim (0.1, -2.0)$ (wide 
solution) and $\sim (-0.1, -2.0)$ (close solution). The lensing parameters of the two solutions 
obtained by refining these locals are listed in Table~\ref{table:three} (standard model) along with 
the $\chi^2$ values of the fits for the individual solutions.  The $M_1$--$M_3$ separations of the 
two solutions are approximately in the relation of $s_{3,{\rm w}}\times s_{3,{\rm c}}\sim 1.0$, 
indicating that the degeneracy between the two solutions is caused by the close-wide degeneracy 
\citep{Griest1998, Dominik1999}. Here $s_{3,{\rm w}}$ and $s_{3,{\rm c}}$ denote the $M_1$--$M_3$ 
separations of the solutions with $s_3>1.0$ and $s_3<1.0$, respectively. It is found that the wide 
solution provides a better fit than the close solution with $\Delta\chi^2=23.1$, which is significant
enough to resolve the degeneracy between the solutions. According to the wide 3L1S solution, the 
mass ratio between $M_1$ and $M_3$, $q_3=9.9\times 10^{-3}$, is very small, indicating that the 
third body is a planet-mass object.  The projected separations of $M_2$ and $M_3$ from $M_1$ for 
the (wide) 3L1S solution, $s_2\sim 1.12$ and $s_3\sim 1.23$, are similar to each other.  According 
to Equation (1) of \citet{Holman1999}, the maximum ratio between $M_3$--$M_1$ and $M_2$--$M_1$ 
separations for the dynamical stability of the planet is $\sim 0.42$ assuming a circular planet 
orbit.  The ratio $s_3/s_2\sim 1.1$ is substantially greater than this critical ratio.  Then, 
$M_2$ should lie at a large separation along the line of sight in front of or at the back of 
$M_1$ in order for the planetary system to be dynamically stable.

The model curves of the wide 3L1S model (solid curve) and the residuals from the model are 
shown in Figure~\ref{fig:two}.  It is found that the 3L1S model well describes all the anomaly 
features, significantly improving the fit, by $\Delta\chi^2=2407$, with respect to the standard 
2L1S model. The fit improvement occurs not only around the region of the first peak, for which 
the 2L1S model exhibited a time offset between the model and data, but also throughout the whole 
anomaly region.

The lens system configuration of the 3L1S solutions is shown in Figure~\ref{fig:six}: upper 
panel for the wide solution and lower panel for the close solution.  The positions of the 
individual lens components are represented by red filled dots, marked by $M_1$, $M_2$, and $M_3$.  
From the comparison of the caustic with the 2L1S caustic, drawn in gray, it is found 
that the tertiary lens component has two effects on the caustic. First, $M_3$ induces a new 
set of caustic at around the source position of $(x_{\rm s}, y_{\rm s}) \sim (0.35, -0.1)$ 
for the wide solution and $\sim (-0.3, -0.3)$ for the close solution, and this makes the 
caustic nested and self-intersecting.  Second, $M_3$ additionally causes a slight distortion 
of the caustic from that of the 2L1S solution. The source did not pass the region near the 
additional caustic structures induced by $M_3$. Nevertheless, the light curve deviates from 
the 2L1S form due to the distortion of the caustic, and this explains the residuals from the 
standard 2L1S model.  If the 3L1S interpretation is correct, KMT-2021-BLG-0322 is the third 
case in which a planet belonging to a binary is detected through the signal from the caustic 
distortion induced by a tertiary lens component after the cases of OGLE-2007-BLG-349 and 
OGLE-2013-BLG-0341.

\begin{figure}[t]
\includegraphics[width=\columnwidth]{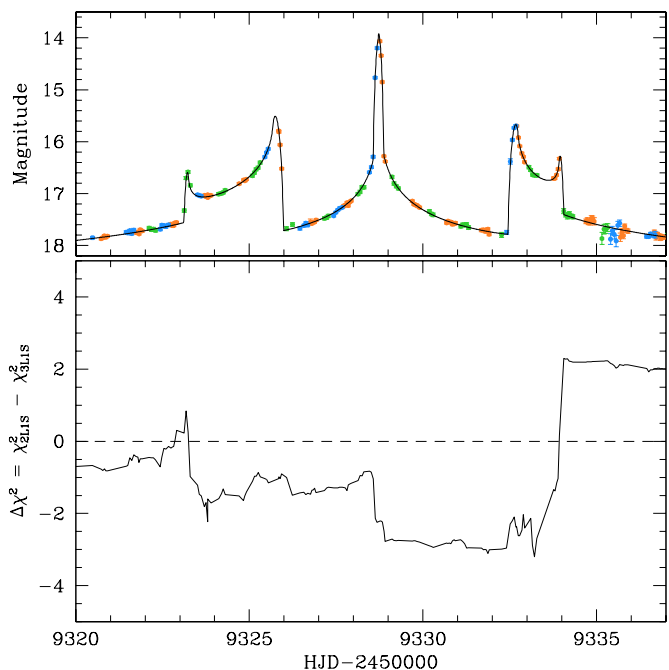}
\caption{
Cumulative distribution of $\chi^2$ difference between higher-order 2L1S and 3L1S models. 
The light curve in the upper panel is inserted to show the region of $\chi^2$ difference.
}
\label{fig:seven}
\end{figure}

\subsection{2L1S versus 3L1S interpretations}\label{sec:three-three}

The analyses conducted in the previous subsections show that the deviations from a standard 
2L1S model in the lensing light curve of KMT-2021-BLG-0322 can be explained almost equally 
well by considering the higher-order effects or by introducing a low-mass tertiary lens 
component.   The degeneracy between the two models can be seen by comparing the residuals 
from the two models presented in Figure~\ref{fig:two}.  In Figure~\ref{fig:seven}, we present 
the cumulative distribution of $\Delta\chi^2 =\chi^2_{\rm 2L1S}-\chi^2_{\rm 3L1S}$  to better 
show the subtle difference between the fits of the two models.  The distribution shows that 
$\Delta\chi^2\lesssim 2$ throughout the major anomaly region, indicating that the degeneracy 
between the two models is very severe.

The degeneracy between a higher-order 2L1S model and a triple-lens model for KMT-2021-BLG-0322 
is very similar to the degeneracies identified in the two previous lensing events MACHO-97-BLG-41 
and OGLE-2013-BLG-0723, for which the residuals from the standard 2L1S models were initially 
interpreted as signals of a planetary-mass companion to a binary lens but later explained by the 
higher-order effects of the 2L1S models.  For these previous events, the degeneracies could be 
resolved because the 2L1S models yielded better fits than the corresponding 3L1S models.  For 
KMT-2021-BLG-0322, on the other hand, the degeneracy between the two interpretations is so severe 
that it cannot be resolved based on only the photometry data.  Therefore, KMT-2021-BLG-0322, 
together with the two previous events, demonstrates that the degeneracy can be not only common 
but also very severe.

The degeneracy also calls into question the idea from \citet{Gould2014} that it would be possible 
to detect planets from the distortions that they induce on the cusps.  This may be possible in some 
cases, but at least in this case, if the planet was the real cause of the light curve discrepancy, 
it could not be distinguished from a nonrectilinear motion.  This emphasizes the need to check 
both interpretations when such deviations are detected in lensing light curves.

\section{Lensing observables}\label{sec:four}

Because it is difficult to single out a model for the observed data, we estimate two sets 
of the physical lens parameters based on the observables of the two possible interpretations 
of the event.  The mass $M$ and distance $\dl$ to the lens can be constrained by measuring 
lensing observables.  The event timescale $\te$ is the first such an observable, and it is 
related to $M$ and $\dl$ by
\begin{equation}
\te = {\thetae \over \mu};\qquad
\thetae = (\kappa M \pi_{\rm rel})^{1/2},
\label{eq3}
\end{equation}
where $\kappa=4G/(c^2{\rm AU})$. The other two observables are the microlens parallax $\pie$ 
and angular Einstein radius $\thetae$, with which $M$ and $\dl$ are uniquely determined by
\begin{equation}
M = {\thetae \over \kappa\pie  };\qquad
\dl = { {\rm AU} \over \pie\thetae+\pi_{\rm S}}, 
\label{eq4}
\end{equation}
where $\pi_{\rm S}={\rm AU}/D_{\rm S}$ is the parallax to the source \citep{Gould2000}.

\begin{figure}[t]
\includegraphics[width=\columnwidth]{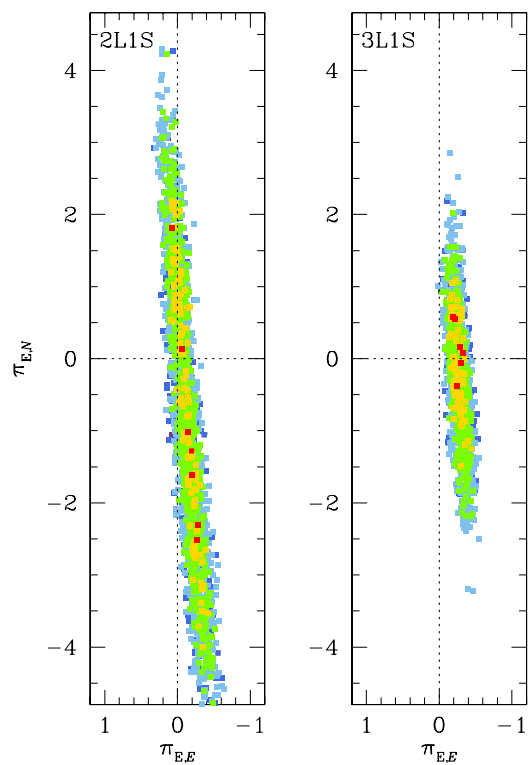}
\caption{
$\Delta\chi^2$ distribution of points in the MCMC chain 
for the higher-order 2L1S (left panel) and 3L1S wide (right panel) solutions
plotted on the $\piee$--$\pien$ plane. 
Color coding is same as that of Fig.~\ref{fig:four}.
}
\label{fig:eight}
\end{figure}

The microlens parallax is measured from the slight deviation in the lensing light curve 
caused by the orbital motion of Earth.  The microlens-parallax parameters of the 2L1S model 
are listed in Table~\ref{table:two}.  For the measurements of the parallax parameters 
corresponding to the 3L1S model, we conduct an additional modeling by considering the 
microlens-parallax effect, and the lensing parameters of the model including $(\pien, \piee)$ 
are listed in Table~\ref{table:three}.  It is found that the consideration of the microlens-parallax 
effect improves the 3L1S fit by $\Delta\chi^2=11.6$ with respect to the standard model.  The left 
and right panels of Figure~\ref{fig:eight} show the $\Delta\chi^2$ plots of MCMC points on the 
$\piee$--$\pien$ plane for the 2L1S and 3L1S solutions, respectively.  For both models, it is 
found that the uncertainty of the north component of the microlens-parallax vector is large, 
while the east component is relatively well constrained.

The angular Einstein radius is measured by analyzing the deviations of the lensing light 
curve from a point-source form caused by finite-source effects during the caustic crossings. 
This analysis yields the normalized source radius $\rho$, and the angular Einstein radius is 
determined by $\thetae = \theta_*/\rho$, where the angular source radius $\theta_*$ is deduced 
from the color and brightness of the source.  We estimate the extinction and reddening-corrected 
(dereddened) source color and brightness, $(V-I, I)_0$, from the instrumental values, $(V-I, I)$, 
using the \citet{Yoo2004} method, in which the centroid of red giant clump (RGC) in the 
color-magnitude diagram (CMD) is used as a reference for this conversion.  The RGC centroid 
can be used as a reference because its dereddened color and brightness, $(V-I, I)_{\rm RGC,0}$, 
are known, and the source and RGC stars, both of which are located in the bulge, experience 
similar reddening and extinction.

\begin{figure}[t]
\includegraphics[width=\columnwidth]{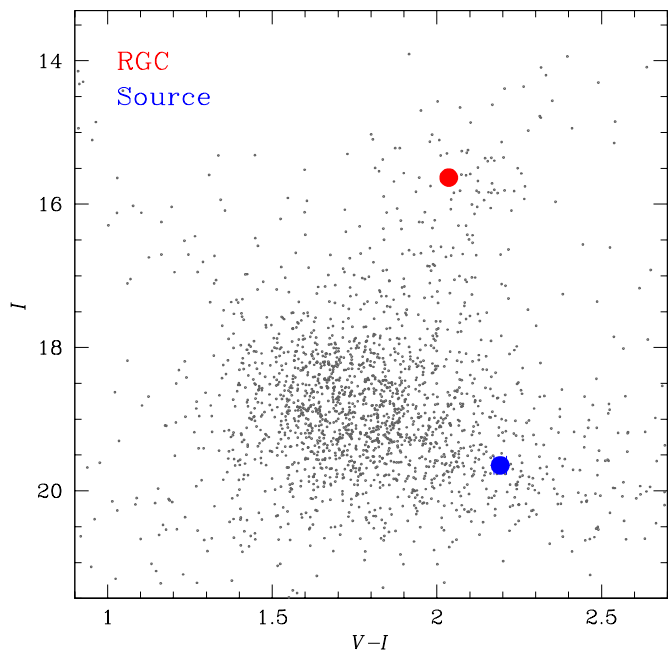}
\caption{
Positions of the source with respect to the centroid of red giant clump (RGC) in the instrumental 
color-magnitude diagram of stars located around the source.
}
\label{fig:nine}
\end{figure}

Figure~\ref{fig:nine} shows the source position (blue dot) with respect to the RGC centroid (red 
dot) in the CMD of stars around the source.  The source color and magnitude are estimated from the 
regression of the KMTC pyDIA data to the model, and align them to the OGLE-III system to show 
calibrated color and magnitude values.  The instrumental colors and magnitudes of the source 
and RGC centroid are $(V-I, I)=(2.19\pm 0.02, 19.64\pm 0.01)$ and $(V-I, I)_{\rm RGC}= (2.04, 
15.55)$, respectively. From the measured offsets in the color and magnitude between the source 
and RGC centroid, $\Delta (V-I, I)$, together with the known dereddened values of the RGC 
centroid $(V-I, I)_{\rm RGC,0}=(1.06, 14.38)$ \citep{Bensby2013, Nataf2013}, the dereddened 
source color and brightness are estimated as
\begin{equation}
\eqalign{
(V-I, I)_0 & =  (V-I, I)_{{\rm RGC},0 }+ \Delta (V-I, I) \cr
           & =  (1.22\pm 0.02, 18.47\pm 0.01).       \cr
}
\label{eq5}
\end{equation}
We note that the source color and brightness are estimated using the higher-order 2L1S and 
the 3L1S model result in consistent values.  The measured $V-I$ color is then converted into 
$V-K$ color using the \citet{Bessell1988} color relation, and we then estimate $\theta_*$ from 
the $(V-K)$--$\theta_*$ relation of \citet{Kervella2004}. This procedure yields the angular 
source radius of
\begin{equation}
\theta_* = 1.15 \pm 0.08~\mu{\rm as}. 
\label{eq6}
\end{equation}
With the measured angular source radius, the angular Einstein radius and the relative 
lens-source proper motion are determined as
\begin{equation}
\thetae = {\theta_* \over \rho} = 0.43 \pm 0.03~{\rm mas},
\label{eq7}
\end{equation}
and
\begin{equation}
\mu = {\thetae\over \te} = 6.18 \pm 0.48~{\rm mas~yr}^{-1}, 
\label{eq8}
\end{equation}
respectively.

\begin{figure}[t]
\includegraphics[width=\columnwidth]{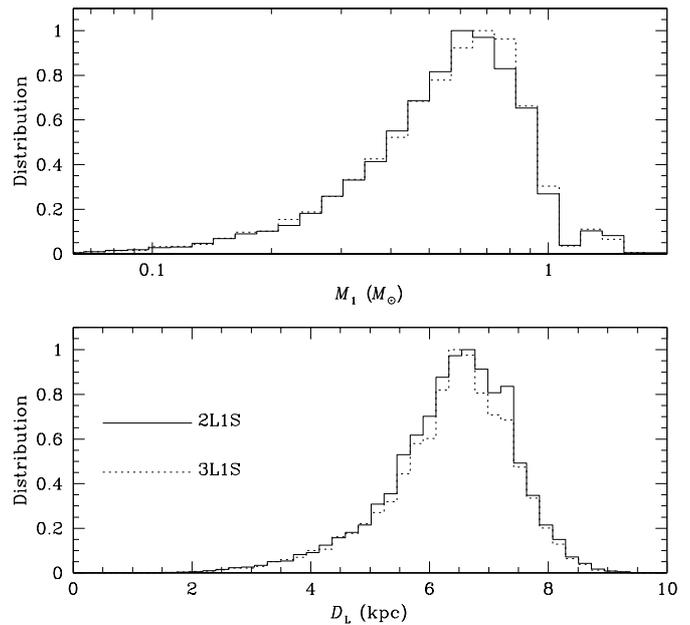}
\caption{
Bayesian posteriors of the primary lens mass ($M_1$,upper panel) and distance (lower 
panel) to the lens.  In each panel, the solid and dotted curves represent the distributions 
obtained using the observables of the 2L1S and 3L1S solutions, respectively.  
}
\label{fig:ten}
\end{figure}

\section{Physical lens parameters}\label{sec:five}

In this section, we estimate the physical parameters of the mass and distance to the lens 
by conducting a Bayesian analysis using the measured observables of $(\te, \theta, \pie)$ 
together with the prior models of the lens mass function and the physical and dynamical 
distributions of Galactic objects.  Because the degeneracy between the 2L1S and 3L1S 
solutions cannot be broken, we conduct two sets analysis based on the parameters and 
observables of the individual solutions.

We carry out the Bayesian analysis in two steps. In the first step, we generate a large number 
($6\times 10^6$) of artificial lensing events by conducting a Monte Carlo simulation, in which 
the mass of the lens, the distance to the lens and source, and the transverse lens-source 
velocity, $v_\perp$, are deduced from the Galactic model. In the Galactic model used in the 
simulation, we adopt the \citet{Han2003}, \citet{Han1995}, and \citet{Zhang2020} models for the 
physical, dynamical distributions, and mass function, respectively. For the individual simulated 
events, we compute the lensing observables using the relations $\te =\dl\thetae/v_\perp$, 
$\thetae =(\kappa M \pi_{\rm rel})^{1/2}$, and $\pie=\pi_{\rm rel}\thetae$. In the second step, 
we construct the posterior distributions of the $M$ and $\dl$ for the artificial events with 
observables that are consistent with the measured values.

Figure~\ref{fig:ten} shows the posterior distributions of the primary lens mass ($M_1$) and 
distance obtained from the Bayesian analysis.  The distributions of the 2L1S (solid curve) and 
3L1S (dotted curve) solutions are similar to each other due to the similarity of the observables 
between the two solutions.  The estimated masses of the binary lens components are
\begin{equation}
M_1=0.62^{+0.25}_{-0.26}~M_\odot;\qquad  
M_2=0.07^{+0.03}_{-0.03}~M_\odot, 
\label{eq9}
\end{equation}
for both 2L1S and 3L1S solutions, indicating that the binary is composed of a K-type dwarf and 
a low-mass object at the star/brown-dwarf boundary.  The mass of the tertiary lens component 
according to the 3L1S solution is 
\begin{equation}
M_3=6.40^{+2.64}_{-2.78}~M_{\rm J},
\label{eq10}
\end{equation}
which is in the planetary mass regime.  If the 3L1S interpretation could be shown to be correct, 
then the planet would be the seventh microlensing planet in a binary.  The system is located at 
a distance from Earth of
\begin{equation}
\dl = 6.6^{+0.9}_{-1.1}~{\rm kpc}.
\label{eq10}
\end{equation}
For the estimation of $M$ and $\dl$, we take the median as representative values and the errors
are estimated as the 16\% and 84\% of the distributions.

\section{Conclusion}\label{sec:six}

We investigated the microlensing event KMT-2021-BLG-0322, for which the lensing light curve 
exhibited three distinctive sets of caustic-crossing features.  It was found that the overall 
feature of the light curve was approximately described by a binary-lens model, but the model 
left substantial residuals.  We tested various interpretations with the aim of explaining the 
residuals.  From this investigation, it was found that the residuals could be explained either 
by considering a nonrectilinear relative lens-source motion caused by the microlens-parallax 
and lens-orbital effects or by introducing a low-mass companion to the binary lens.  The 
degeneracy between the higher-order 2L1S model and the triple-lens model was very severe, 
making it difficult to single out a correct solution  based on only the photometric data.  
This degeneracy was known before for two previous events (MACHO-97-BLG-41 and OGLE-2013-BLG-0723), 
which led to the false detections of planets in binary systems.  Therefore, KMT-2021-BLG-0322, 
together with the two previous events, demonstrates that the degeneracy can be not only common 
but also very severe, emphasizing the need to check both models in the interpretations of 
deviations from 2L1S models.  From the Bayesian analysis conducted with the measured lensing 
observables, it was estimated that the binary lens components have masses $(M_1, M_2) =
(0.62^{+0.25}_{-0.26}~M_\odot,0.07^{+0.03}_{-0.03}~M_\odot)$, for both 2L1S and 3L1S solutions, 
and the mass of the tertiary lens component according to the 3L1S solution is 
$M_3=6.40^{+2.64}_{-2.78}~M_{\rm J}$.

\begin{acknowledgements}
Work by C.H. was supported by the grants  of National Research Foundation of Korea 
(2020R1A4A2002885 and 2019R1A2C2085965).
This research has made use of the KMTNet system operated by the Korea Astronomy and Space 
Science Institute (KASI) and the data were obtained at three host sites of CTIO in Chile, 
SAAO in South Africa, and SSO in Australia.
Work by I.K. was supported by JSPS KAKENHI Grant Number 20J20633.  Work by D.P.B., A.B., 
and C.R. was supported by NASA through grant NASA-80NSSC18K027.  T.S. acknowledges the 
financial support from the JSPS, JSPS23103002, JSPS24253004, and JSPS2624702.  Work by N.K.  
is supported by JSPS KAKENHI Grant Number JP18J00897.  The MOA project is supported by J
SPS KAK-ENHI Grant Number JSPS24253004, JSPS26247023, JSPS23340064, JSPS15H00781, JP16H06287, 
17H02871, and 19KK0082.
\end{acknowledgements}

\end{document}